\documentclass[a4paper]{article}
\usepackage[english]{babel}
\usepackage[utf8x]{inputenc}
\usepackage[T1]{fontenc}
\usepackage[a4paper,top=3cm,bottom=2cm,left=3cm,right=3cm,marginparwidth=1.75cm]{geometry}
\usepackage{amsmath}
\usepackage{graphicx}
\usepackage[colorinlistoftodos]{todonotes}
\usepackage[colorlinks=true, allcolors=blue]{hyperref}
\usepackage{hyperref}
\usepackage{authblk}
\usepackage{epstopdf}
\usepackage{mathtools}

\DeclarePairedDelimiter\floor{\lfloor}{\rfloor}

\begin{document}
\title{Toward Stronger Robustness of Network Controllability:\\A Snapback Network Model %
	\footnote{This research was supported by the Hong Kong Research Grants Council under the GRF Grant 11200317, the National Natural Science Foundation of China under Grant No. 61473189, and the Natural Science Foundation of Shanghai under Grant No. 17ZR1445200.}
    \footnotetext{\llap{\textsuperscript{*}}Corresponding author: Guanrong Chen}}%

\author[1]{Yang Lou}
\author[2,3]{Lin Wang}
\author[1]{Guanrong Chen}
\affil[1]{\textit{Department of Electronic Engineering, City University of Hong Kong, Hong Kong SAR, China}}
\affil[2]{\textit{Department of Automation, Shanghai Jiao Tong University, Shanghai, China}}
\affil[3]{\textit{Key Laboratory of System Control and Information Processing, Ministry of Education, Shanghai, China}}
\maketitle

\begin{abstract}
\normalsize{A new complex network model, called \textit{q}-snapback network, is introduced. Basic topological characteristics of the network, such as degree distribution, average path length, clustering coefficient and Pearson correlation coefficient, are evaluated. The typical 4-motifs of the network are simulated. The robustness of both state and structural controllabilities of the network against targeted and random node- and edge-removal attacks, with comparisons to the multiplex congruence network and the generic scale-free network, are presented. It is shown that the \textit{q}-snapback network has the strongest robustness of controllabilities due to its advantageous inherent structure with many chain- and loop-motifs.}\\
\end{abstract}


\section{Introduction}\label{sec:intro}

The subject of complex networks has gained popularity after two decades of research pursuits
with great efforts from various scientific and engineering communities through intensive and extensive studies,
and has literally become a self-contained discipline interconnecting network science, systems engineering,
statistical physics, applied mathematics and social sciences \cite{Barabasi2016NS,Newman2010N,Chen2014FCN}.

This interdisciplinary research area, the interaction between network science and control systems theory in
particular, has seen very rapid growth since year 2002 \cite{Liu2016RMP,Chen2014IJCAS,Li2004TCS,Wang2002PA}.
In fact, it has created a corpus of new opportunities and yet also great challenges for classical control and
systems theories and technologies, since a complex dynamical network typically has large numbers of nodes and
edges, with higher-dimensional dynamical node-systems interconnected in a complicated structure such as random,
small-world or scale-free topology. For a complex dynamical network, to achieve an optimal objective, practically
one can only control a small fraction of nodes and/or edges via external inputs. These observation and demand
had motivated the long-term endeavor and development of the so-called ``pinning control'' strategy \cite{Wang2002PA},
as a practical control approach to addressing the fundamental questions of how many and which nodes to pin
(to control), aiming to design effective control algorithms that could ``pull one hair to move the whole body''.

For a single-input/single-output (SISO) connected and directed framework of linear time-invariant (LTI)
node-systems, the minimum number of external inputs (controllers) required for the network to be structurally
controllable is determined by a criterion based on maximum matching. The ``minimum inputs theorem'' \cite{Liu2011N}
states that, if a network of size $N$ has a perfect matching then the number of external controllers is $N_D=1$
and the controller can be pinned at any node; otherwise, $N_D=N-|E^*|$ , where $|E^*|$ is the number of elements
in a maximum matching $E^*$, and the controllers should be pinned at the unmatched nodes.

In the multi-input/multi-output (MIMO) setting, the state controllability of a connected, directed and weighted
network of LTI node-systems were studied in \cite{Wang2016A,Wang2017RPT}, where it shows how the network
topology, node dynamics, external control inputs, and inner interactions altogether affect the state
controllability of the network, with necessary and sufficient conditions derived for the controllability of
a connected, directed and weighted network in a general topology. In \cite{Wang2016A,Wang2017RPT}, precise
necessary and sufficient conditions are given in terms of the node-system matrices, control gains and the
network connectivity matrices. Then, some easily-verified formulas were derived in \cite{Hou2017IJRNC} for
verifying the necessary and sufficient controllability conditions for networked MIMO LTI node-systems.
Moreover, both state and structural controllabilities for temporal networks, namely the about MIMO LTI
setting with a certain time-varying network structure, were studied in \cite{Hou2016TCS}, establishing
some necessary and/or sufficient conditions for the network controllability. Furthermore, in \cite{Hou2017TCNS},
conditions and methods were developed for designing the control input matrices of pinning controllers to
guarantee the network state and structural controllabilities.

In retrospect, there had been significant progress in the studies of network controllability in the past decade
\cite{Liu2016RMP}-\cite{Tang2017AXV}
These studies were concerned with, for example,
pinning small unmatched nodes \cite{Yan2012PRL}, optimizing the network controllability \cite{Wang2012PRE},
identifying critical nodes for controllability \cite{Jia2013NC},
investigating the exact controllability \cite{Yuan2013NC}, finding the importance of in- and out-degrees
for control \cite{Menichetti2014PRL}, and designing targeted control \cite{Gao2014NC}. More recently,
studies have evolved to considering, for instance, mathematical and computational approaches to
controlling nonlinear networks \cite{Motter2015C}, control properties of complex networks \cite{Ruths2015S},
control energy issue \cite{Yan2015NP},
sensor-actuator placements for network controllability \cite{Summers2016TCNS},
human protein-protein interaction network \cite{Vinayagam2016PNAS}, structural controllability of
temporal networks \cite{Yao2017PO}, turning physically uncontrollable networks to become controllable
ones \cite{Wang2017SR},
and so on. Other related works on network controllability can be found from the recent review articles
 \cite{Liu2016RMP}, \cite{Chen2017}.

On the other hand, the issue of network robustness has been extensively investigated by different means
under different criteria in different settings, and there is a vast volume of literature on the subject.
Concerning the robustness of the controllability of a complex network against node and/or edge removals,
which typically cause cascading failures \cite{Buldyrev2010N,Motter2002PRE}, so that the network could
retain its connectivity and functionality (here, the network controllability), relevant research includes
the following. In \cite{Mishkovski2011CNSNS}, the normalized average edge betweenness is used as a measure
of the network vulnerability; in \cite{Zeng2012PRE}, a hybrid method combining similarity-based index and
edge-betweenness centrality is proposed for identifying and removing spurious interactions, keeping the
network connectivity and preserving the network functionality; in \cite{Pu2015PA}, it investigates the
vulnerability of complex networks subject to path-based attacks, showing that the more homogeneous the
degree distribution is, the more fragile the network will be. Particularly related to the present concern
on the network controllability, in \cite{Lu2016PO} the vulnerability of network controllability is
considered, where the attacks are based on node degrees or edge betweenness, with simulations showing
that the node-based attacks are more harmful to the network controllability than the edge-based attacks
and that heterogeneous networks are more vulnerable than homogeneous ones; it was found, however, that
for many real-world networks the betweenness-based attacks are actually most harmful to the network
controllability.

A piece of recent theoretical work on network controllability and the corresponding robustness is the
initiation of a mathematical number-theoretic framework of complex network modeling and analysis
 \cite{Yan2016SR}. A congruence network is generated as follows. A link (edge) in the congruence network
is defined according to the congruence relation $j\equiv r$ (mod $i$), where $r$ is the reminder of $j$
divided by $i$, and they are all integers (here, natural numbers). For every fixed $r$, an infinite set
of natural number pairs ($i$, $j$) can be generated. For each pair of such integers, a directed link from
$i$ to $j$ ($i<j$) characterizes the congruence relation between them. For each $r$, this process yields
a congruence network associated with the reminder $r$, denote by $G$($r$, $N$), where $N$ is the largest
natural number in the present construction of the network. Then, for different values of $r$ ($r\le N$),
one obtains various such networks, referred to as multiplex congruence networks (MCN). It was found that
every MCN is precisely a scale-free network with a power-law distribution for out-degrees \cite{Yan2016SR}.

From the construction of an MCN, one can see that it contains many chains, where as usual every
chain has a root-node. Therefore, for each chain, using one external linear self-state feedback controller
is sufficient to guarantees the controllability of the chain. Since typically $r \ll N$, the controllability
of the entire network is excellent, in the sense that a very small number of controllers can guarantee
the controllability of a large network. This is quite opposite to the common view that scale-free networks
are generally not good in controllability by requiring large numbers of controllers because many small
nodes need to be individually controlled in general. Moreover, when a chain is being attacked, randomly
or intentionally, with one node or one edge removed, in the worst situation it is broken into two
sub-chains. In this case, at most one new controller would need to be added at the new chain-root in
order to retain the controllability of the network, so it is very robust against attacks. This is also
quite opposite to the common view that a scale-free network is fragile against intentional attacks.

The above interesting findings have stimulated our curiosity about the network controllability and its
robustness against attacks, urging us to find out why, how, and what the key factors are behind the
surprising phenomena regarding the controllability and its robustness against malicious attacks for
general complex dynamical networks. In this paper, we attempt to modify and extend the multi-chain
structure of the MCN to a multi-ring structure, thereby proposing a new \textit{q}-snapback network
model, which will be shown to be superb in the robustness of network controllability. Extensive
simulation results indeed demonstrate that \textit{q}-snapback networks and MCN outperform general
scale-free networks in resisting both targeted and random attacks, and also demonstrate that
the \textit{q}-snapback network is prominently more robust than the MCN against targeted attacks
on the nodes with largest betweenness, and also against random attacks. The \textit{q}-snapback network
has similar robustness as the MCN when the targeted attack aims at removing highest degree nodes.

The main contribution of this paper is the introduction of a new network model based on
the novel idea of using snapback connections, which turned out to be a good model with the strongest
robustness of network controllability against both targeted and random node/edge removal attacks.
One technical challenge was to reveal and confirm the key network sub-structures that affect the
controllability robustness of the new network model, which were found to be the relatively large
numbers of chains and loops existing in the new model, which are not prominent in other well-known
network models.

The rest of the paper is organized as follows. Section~\ref{sec:model} describes the \textit{q}-snapback
network model. Section~\ref{sec:anl} presents some analysis on the degree distribution of the new model.
Section~\ref{sec:sim} shows simulation results on various topological features especially degree
distributions and motifs. Section~\ref{sec:ctrl} discusses both state and structural controllabilities
and compares their robustness for three types of networks, \textit{i.e.}, MCN, scale-free
and \textit{q}-snapback networks. Section~\ref{sec:con} concludes the investigation.\\

\section{The \textit{q}-snapback Network Model}\label{sec:model}

\noindent
Although both the MCN and generic scale-free networks have power-law degree distributions, they behave
oppositely against targeted and random attacks; namely, as is well known, generic scale-free networks
are robust against random attacks but fragile against targeted attacks, but in contrast MCNs are robust
against targeted but fragile against random attacks \cite{Yan2016SR}. This suggests that the power-law
degree distribution is not the essential reason supporting the network robustness against attacks and
failures.

It is observed that an MCN contains many chains. Since each chain has one root, a subgraph with $r$
chains has at most $r$ roots. According to the matching theory \cite{Liu2011N}, to control a chain
only one controller is needed to pin at the root. The chain structure is robust against targeted attack
regarding the network controllability, since after one node-removal at most one more new controller is
needed to retain its controllability.

\begin{figure}[t!]
\begin{center}
\includegraphics[width=3.5in]{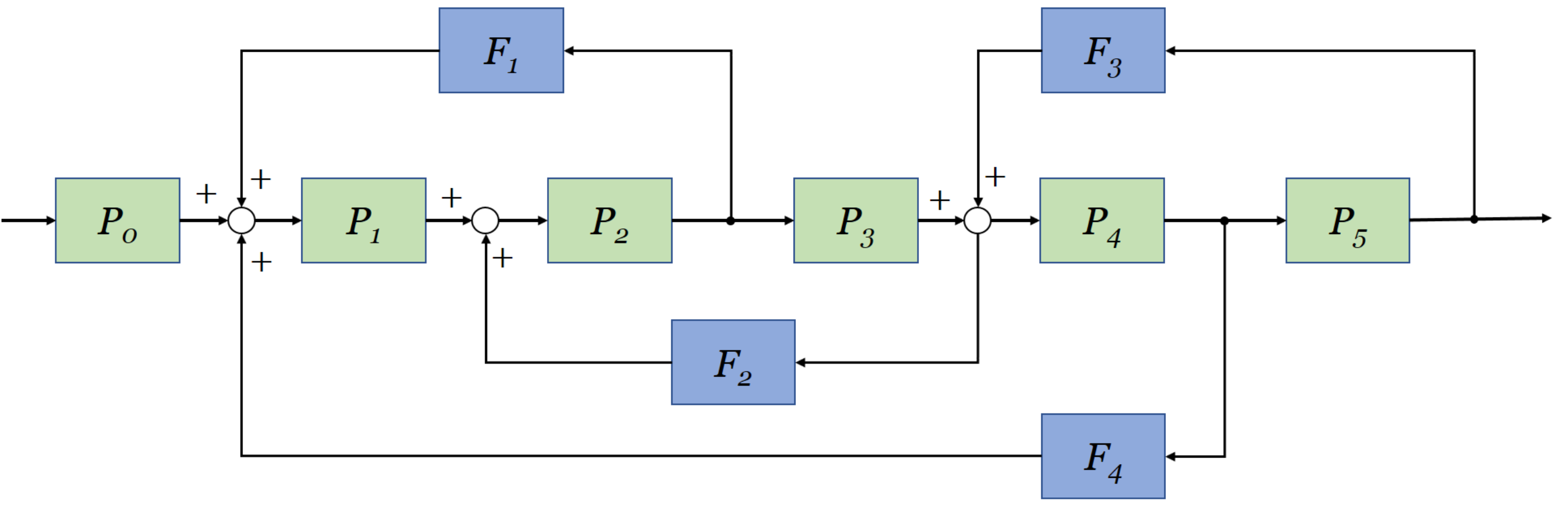}
\caption{An example of assembly-line automation with a snapback connection structure, where
$P_i$ represents the $i$th plant and $F_j$ represents the $j$th feedback controller.}\label{fig1_eg}
\end{center}
\end{figure}

It is also observed that, although the structure of chains offers a good controllability to MCN, these
chains have feedforward loop connections. Practically, feedback loops are more common and more useful
than feedforward ones. For example, the industrial assembly-line illustrated by {Fig.~\ref{fig1_eg} is
very common in manufacturing processes. On the other hand, the number-theoretic congruence relation has
no patterns and no analytic formulas to use for design and analysis considerations. Therefore, a model
with feedback connections (called \textit{snapback~links}) based on a uniform probability distribution
is proposed.

In the snapback network model: 1) the feedforward links are replaced by feedback links;
2) the congruence relations are replaced by uniform random connections. Similarly to the MCN, the basic
structure of the new model is a backbone chain, which is a maximum matching with all matched nodes
except the root. The main difference of the new model compared with the MCN model lies
in its large number of loops, which turns out to be advantageous for the controllability robustness
as further discussed later below.

More precisely, the \textit{q}-snapback network consists of multiple layers, generated as follows. Let
$q\in[0,1]$ be a probability parameter. Each layer starts with a directed chain, which will be the
backbone of the connected layer. Then, following some rules a number of snapback links are generated
connecting to the chain with probability \textit{q}. Finally, all layers are stacked together as a
whole network, where the same nodes will merge into a single node and the same links will also merge
into a single link, so as to avoid multiple nodes and multiple links.

Specifically, start from a directed chain with nodes $1, 2, ..., N$.

For $r=1$, process as follows to generate the first layer network. For every node $i=2,3,4,...,N$,
it connects backward to all previously-appeared nodes $i-1,i-2,...,2,1$, all with a (same and small)
probability $q\in[0,1]$. As a result, some will be backward connected but some will not, which
happens at random uniformly.

For $r=2$, continue to process in the same way but it connects backward to only some (not all)
previously-appeared nodes, \textit{i.e.}, for nodes $i=3,4,...,N$, they backward connect to nodes
$i-2,i-4,...,i-2\floor{\frac{i}{2}}$, with the same probability $q\in[0,1]$ uniformly. In notation,
if $i-2\floor{\frac{i}{2}}=0$, then the link $(i,i-2\floor{\frac{i}{2}})$ will not exist.

Next, the construction continues similarly. For the $r$th layer of the network, with $r=3,4,5,...,N-1$,
the nodes $i=r+1,r+2,...,N$ will backward connect to nodes $i-r,i-2r,...,i-r\floor{\frac{i}{r}}$,
with the same probability $q\in[0,1]$ uniformly. Denoted it as $G_r(q,N)$.

The above procedure continues, until it cannot be processed any further.

Finally, stack all so-generated layers together into one, thus establishing the final multiplex network,
denoted as $G(q,N)=\bigcup_{r=1}^{N-1}G_r(q,N)$, called the \textit{q}-snapback multiplex network of
size $N$.

\begin{figure}[htbp]
\begin{center}
\includegraphics[width=3.5in]{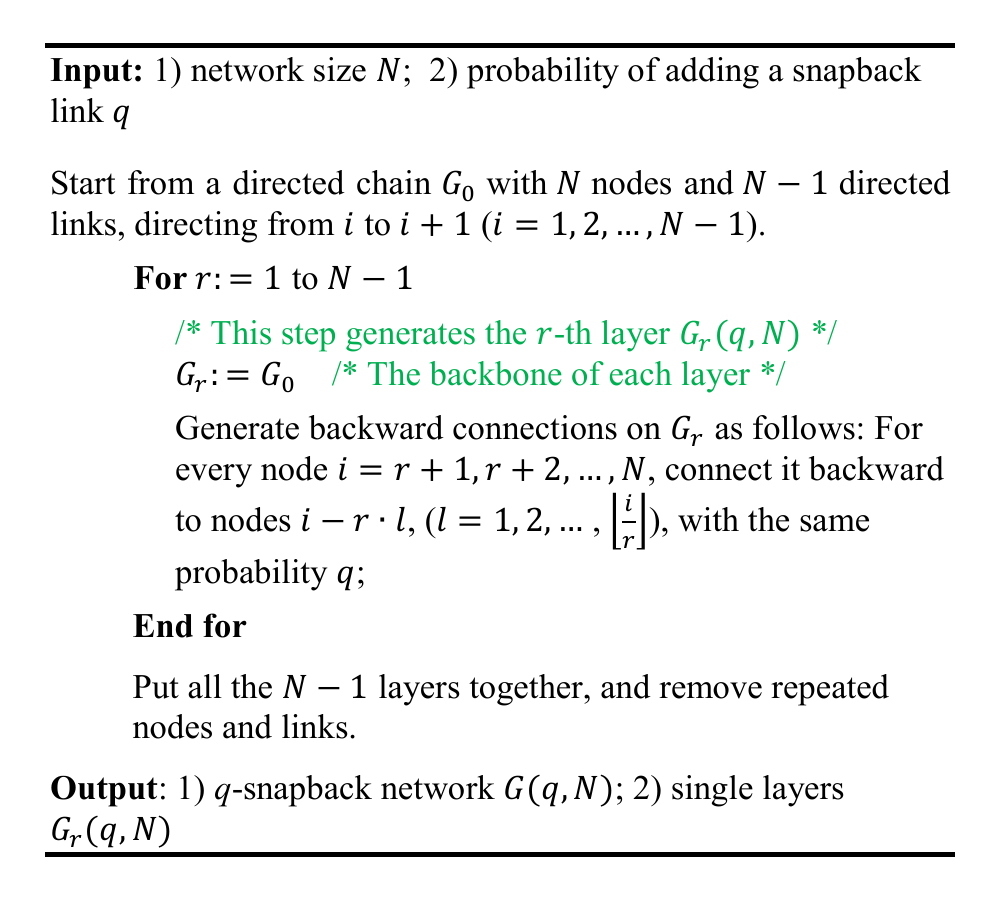}
\caption{Pseudo codes for generating a \textit{q}-snapback multiplex network.}\label{fig2_code}
\end{center}
\end{figure}

Fig.~\ref{fig2_code} shows the pseudo codes for generating a \textit{q}-snapback network. The input
parameters include the network size $N$ and the probability of adding snapback links, $q\in[0,1]$.
Note that, with $q=0$, it is the original chain without any backward connection; with $q=1$, it
is a maximum-size snapback network having the largest number of backward connections.

For each layer, there is a directed chain with some numbers of backward connections. As described
above, with $r=1$, the backward connections on layer $G_1$ are generated as follows: For every
node $i=2,3,4,...,N$, connect it backward to previously-existing nodes $i-1,i-2,i-3,...,2,1$,
all with the same probability $q$. For this case of $r=1$, the generated layer is the densest
one, with the largest number of links. On the contrary, with $r=N-1$, there will be only one
possible backward connection on the chain, \textit{i.e.}, only node $i=N$ could possibly be
connected backward to the first node, which is the sparsest layer (the (\textit{N}$-$1)st layer).

Each layer can work separately and independently, since it is built on the backbone. Also,
all the layers can be stacked together so as to form a multiplex network. Note that all the
repeated nodes and links are removed when the layers are put together as one whole network,
avoiding multiple nodes and links. Fig.~\ref{fig3_layer} shows an example of stacking two
layers together, which has two types of repeated links: 1) the backbone directed links
from $i$ to $i+1$, for $i=1,2,...,N-1$, and 2) the repeated backward links. As can be seen
from the figure, the resultant \textit{q}-snapback network is an ensemble of links from the
two layers without multiple nodes and links.\\

\begin{figure}[t!]
\begin{center}
\includegraphics[width=3.3in]{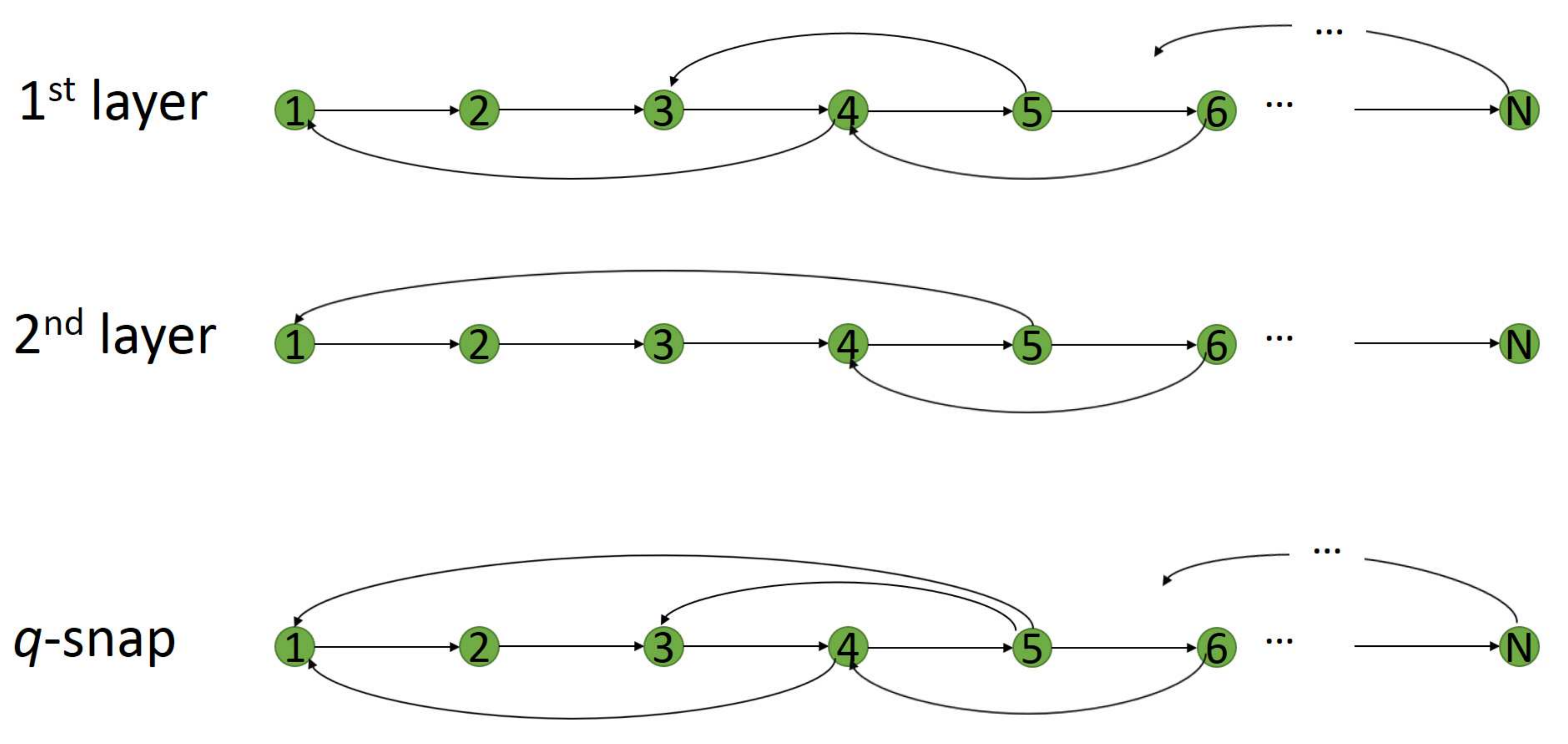}
\caption{A two-layer example of the \textit{q}-snapback multiplex network.
Each layer is a connected subnetwork and their backbone structures are the same directed chain.
On both the 1st and 2nd layers, there is a link connecting from node 6 back to node 4.
When stacking them together in the integrated multiplex network, only one link from node 6
to node 4 is kept. The situation for the other nodes is similar.}\label{fig3_layer}
\end{center}
\end{figure}

\section{Analysis on Degree Distributions}\label{sec:anl}

\noindent In this section, degree distribution of the \textit{q}-snapback network is derived
analytically. Here, the out-degrees and in-degrees of a directed network are both considered.
First, the degree distribution of each single layer is discussed, followed by the multiplex
network.

\subsection{Single layers}

\noindent For the \textit{r}th layer, $r=1,\cdots,N-1$, the out-degree of the \textit{i}th
node $d_O(i)$, $i=1,2,...,N$, is calculated by
\begin{equation}\label{eq1}
d_O(i) =
\begin{cases}
    1,								& \textrm{for}~i=1,2,...,r      \\
    1+\floor{\frac{i-1}{r}}\cdot q,	& \textrm{for}~i=r+1,...,N-1    \\
    \floor{\frac{i-1}{r}}\cdot q,	& \textrm{for}~i=N
\end{cases}
\end{equation}
where $\floor{x}$ is the floor function that returns the greatest integer less than or equal to $x$.

Similarly, the in-degree of the \textit{i}th node $d_{I}(i)$, $i=1,2,...,N$, is
\begin{equation}\label{eq2}
d_I(i) =
\begin{cases}
     \floor{\frac{N-1}{r}}\cdot q,	& \textrm{for}~i=1 			\\
    1+\floor{\frac{N-i}{r}}\cdot q,	& \textrm{for}~i=2,...,N-r	\\
    1,								& \textrm{for}~i=N-r+1,...,N
\end{cases}
\end{equation}
where $\floor{x}$ is the floor function.

\subsection {The multiplex \textit{q}-snapback network}

\noindent When a number of layers are stacked together, the multiplex network is formed. As shown
in Fig.~\ref{fig4_eg}, for any node \textit{i}, its out-degree is
\begin{equation}\label{eq3}
d^M_O(i) =
\begin{cases}
    1,							& \textrm{for}~i=1      \\
    1+\sum_{j=1}^{i-1}{I_{i,j}}q,	& \textrm{for}~i=2,3,...,N-1    \\
    \sum_{j=1}^{N-1}{I_{i,j}}q,		& \textrm{for}~i=N
\end{cases}
\end{equation}
where
\begin{equation}\label{eq4}
I_{i,j}=
\begin{cases}
    1,  & \textrm{if there exists an edge}~(i,j) \\
    0,  & \textrm{otherwise}
\end{cases}
\end{equation}

Similarly, the in-degree of the \textit{i}th node is
\begin{equation}\label{eq5}
d^M_I(i) =
\begin{cases}
	\sum_{j=2}^{N}{I_{i,j}}q,		& \textrm{for}~i=1				\\
    1+\sum_{j=i+1}^{N}{I_{i,j}}q,   & \textrm{for}~i=2,3,...,N-1	\\
    1							& \textrm{for}~i=N
\end{cases}
\end{equation}
where $I_{i,j}$ is defined in (\ref{eq4}).

\begin{figure}[t!]
\begin{center}
\includegraphics[width=2.9in]{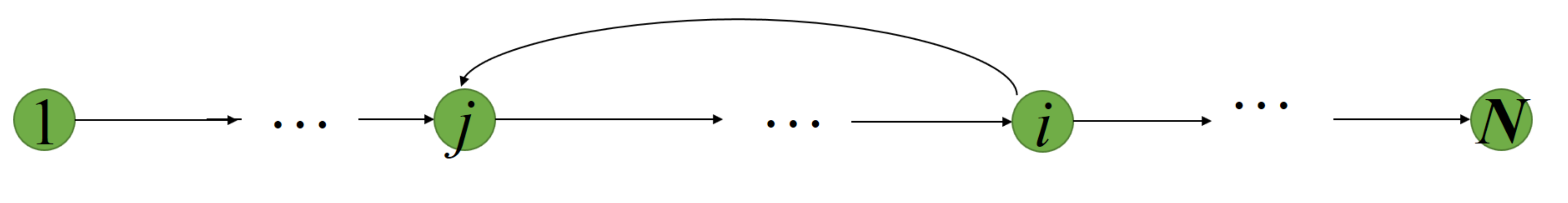}
\caption{Illustration of a snapback link from any node $i$ to any node $j$
($j<i$) in a \textit{q}-snapback network.}\label{fig4_eg}
\end{center}
\end{figure}

An edge $(i,j)$ could appear on different layers, as illustrated in Fig.~\ref{fig4_eg}.
The probability that the edge $(i,j)$ exists on at least one layer, \textit{i.e.},
the probability of existence of edge $(i,j)$, is given by
\begin{equation}\label{eq6}
P_{(i,j)}=1-(1-q)^{^{\prod_{i=1}^{m}{(x_i+1)}}}
\end{equation}
where $i-j=a_1^{x_1}\cdot$$a_2^{x_2}\ldots $$a_m^{x_m}$, with $a_1,a_2,...,a_m$ being prime
numbers, and $(1-q)^{\prod_{i=1}^{m}{(x_i+1)}}$ represents the probability that edge $(i,j)$
does not exist on any layer of the multiplex network. Here, $1$ means $1^0$.\\

\section{Simulations}\label{sec:sim}

\noindent Extensive simulations had been performed on the \textit{q}-snapback network of size
$N=10^{4}$, with $q=0.1$ unless otherwise indicated. The following statistical results are
averages over 50 independent runs.

It was found that:

1) the average path length is 1667.6, with a standard deviation 0.0;

2) the clustering coefficient is 0.4679, with a standard deviation $6.4\times10^{-5}$;

3) the Pearson correlation coefficient (\textit{i.e.}, assortativity) is $-$0.4979, with a
standard deviation $1.1\times10^{-4}$.

Next, the out-degree distributions of some single layers $G_r$($q=0.1$, $N=10^{4}$), and of
the multiplex network $G$($q=0.1$, $N=10^{4}$), are simulated. Here, only the simulation
results on the out-degree distributions are shown, for brevity, since the in-degree
distributions (\ref{eq2}) and (\ref{eq5}) have similar forms as the out-degree ones (\ref{eq1})
and (\ref{eq3}). Moreover, the influence of the parameter \textit{q} on the degree distribution
of the multiplex network is shown and analyzed. Finally, the distribution of the 4-motifs on
the multiplex network is simulated and discussed.

\subsection{Out-degree distributions of single layers}

\noindent Fig.~\ref{fig5_k9} shows the degree distribution of 9 single layers of a multiplex
 \textit{q}-snapback network. For each single layer, the degree distribution is uniform.
The tails of the distribution curves in the figures are due to the well-known finite-size
effects. Both empirical and analytical results are presented in the figures. The empirical
simulation results (presented by blue pluses in the figures) are averages over 50 independent
runs, while the red lines are the analytical solutions calculated by equation (\ref{eq1}) for
reference.

\begin{figure}[t!]
\begin{center}
\includegraphics[width=5.0in]{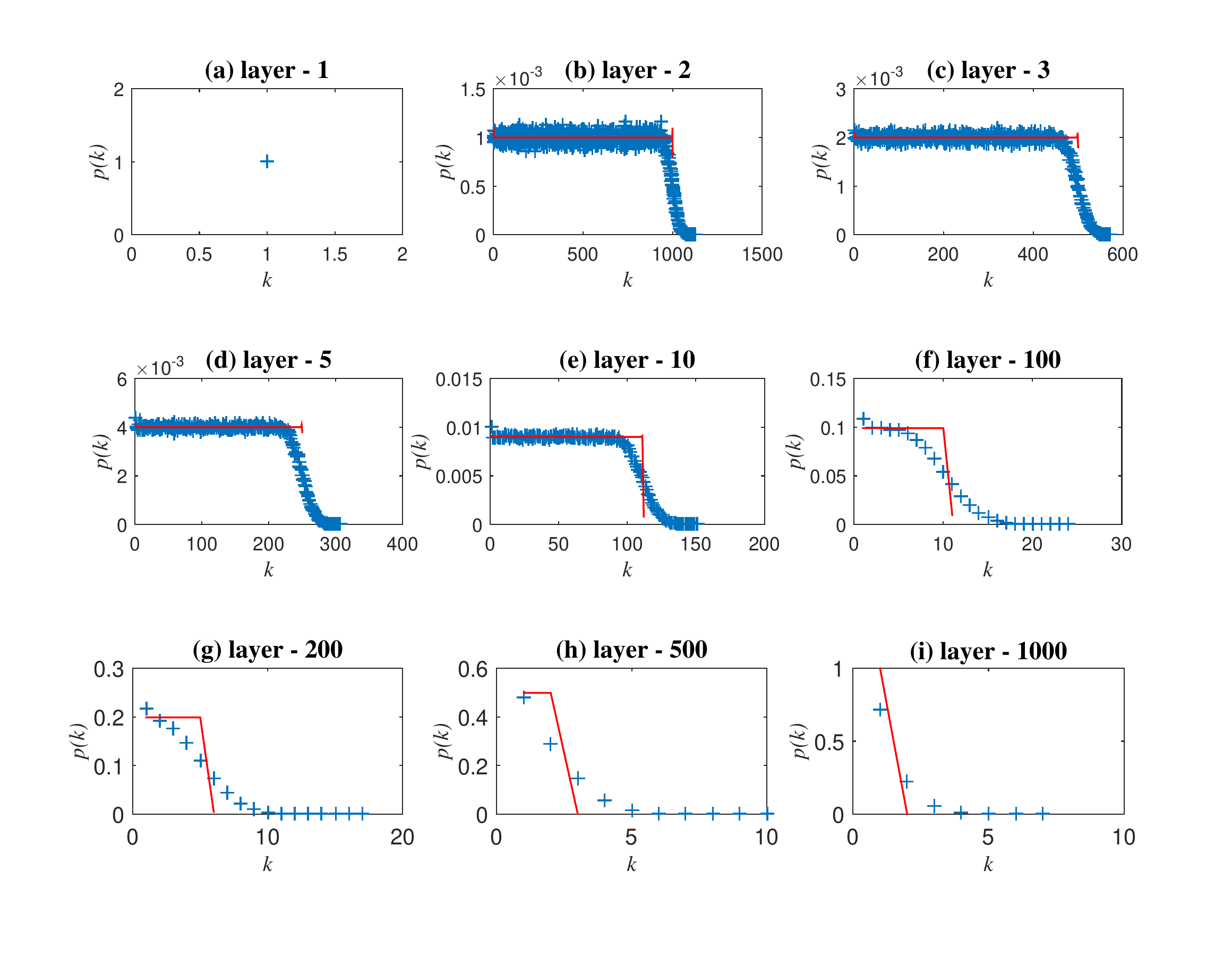}
\caption{Degree distributions of different single layers. The blue pluses (+) are simulation
results and the red lines are calculated using equation (\ref{eq1}): (a) layer $r=1$; (b) layer
$r=2$; (c) layer $r=3$; (d) layer $r=5$; (e) layer $r=10$; (f) layer $r=100$; (g) layer $r=200$;
(h) layer $r=500$; (i) layer $r=1,000$.}\label{fig5_k9}
\end{center}
\end{figure}

\subsection{Out-degree distribution of the multiplex network}

\noindent Fig.~\ref{fig6_k} shows the degree distribution of the multiplex \textit{q}-snapback
network. As can be seen from the figure, the degree distribution is uniform, just like all the
single layers, as expected. The analytical degree distribution calculated by equation (\ref{eq3})
is also plotted in Fig.~\ref{fig6_k}, for reference. Note that equation (\ref{eq3}) gives the
expectation of the out-degrees, which is a real number. When plotting Fig.~\ref{fig6_k}, for better
visualization the real numbers are rounded to their nearest integers, thus the analytical degree
distribution curve appears to be three parallel lines.

\begin{figure}[htbp]
\begin{center}
\includegraphics[width=2.9in]{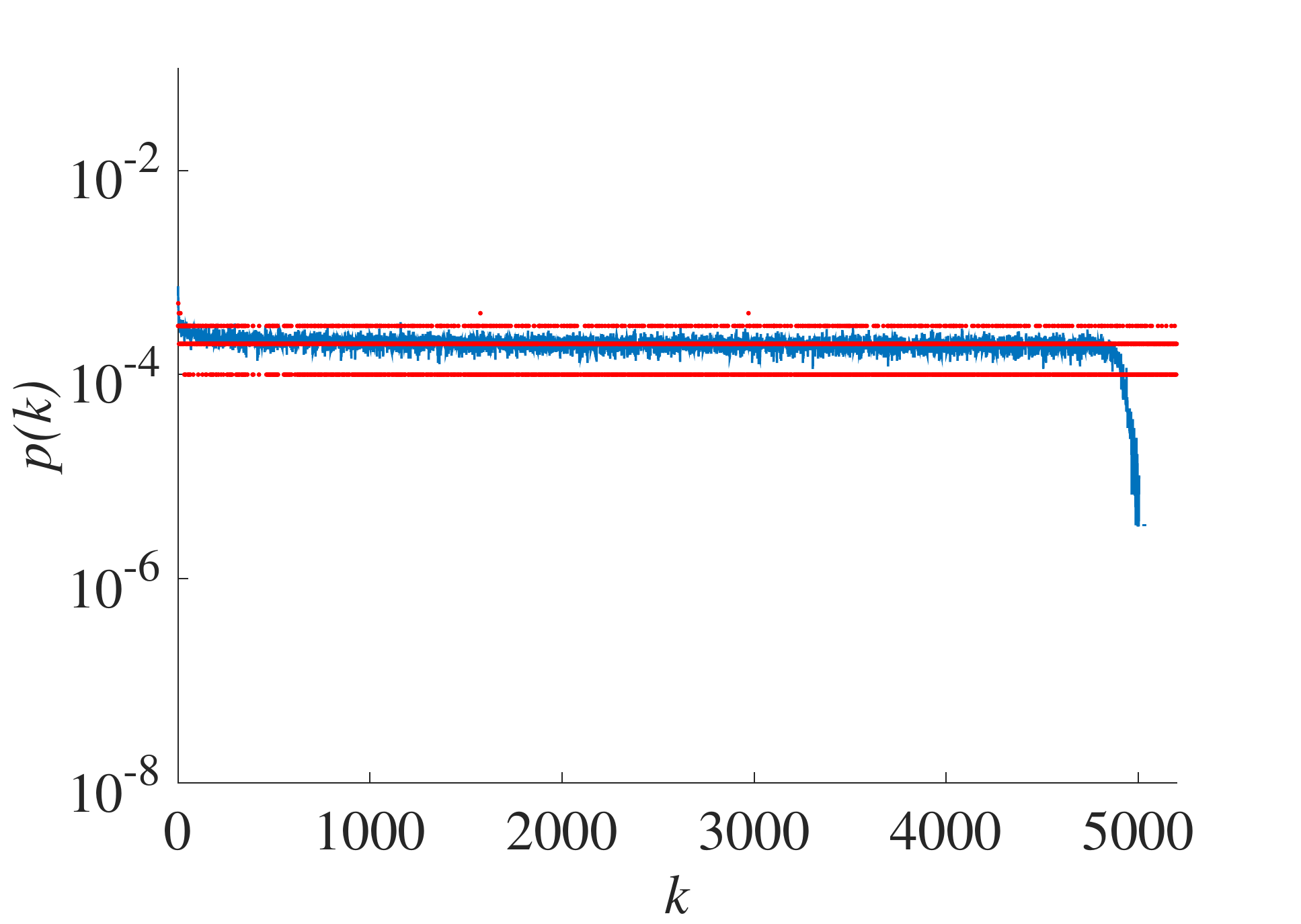}
\caption{The degree distribution of the multiplex \textit{q}-snapback network. The blue pluses
(+) are simulation results and the red dots are calculated using equation (\ref{eq3}), where
real numbers are rounded to their nearest integers thus the analytic curve appears to be
three parallel lines.}\label{fig6_k}
\end{center}
\end{figure}

\subsection{Influence of probability \textit{q} on degree distributions}

\noindent Fig.~\ref{fig7_kvsq} shows the influence of the probability parameter \textit{q} on the degree
distribution of the multiplex network. As can be seen from the figure, the curve becomes more widely
distributed as \textit{q} increases, but it still remains being uniform constantly.
This means that uniform distribution is a scale-free property (bigger value of $q$ generates
more edges) of the new model.

\begin{figure}[t!]
\begin{center}
\includegraphics[width=2.9in]{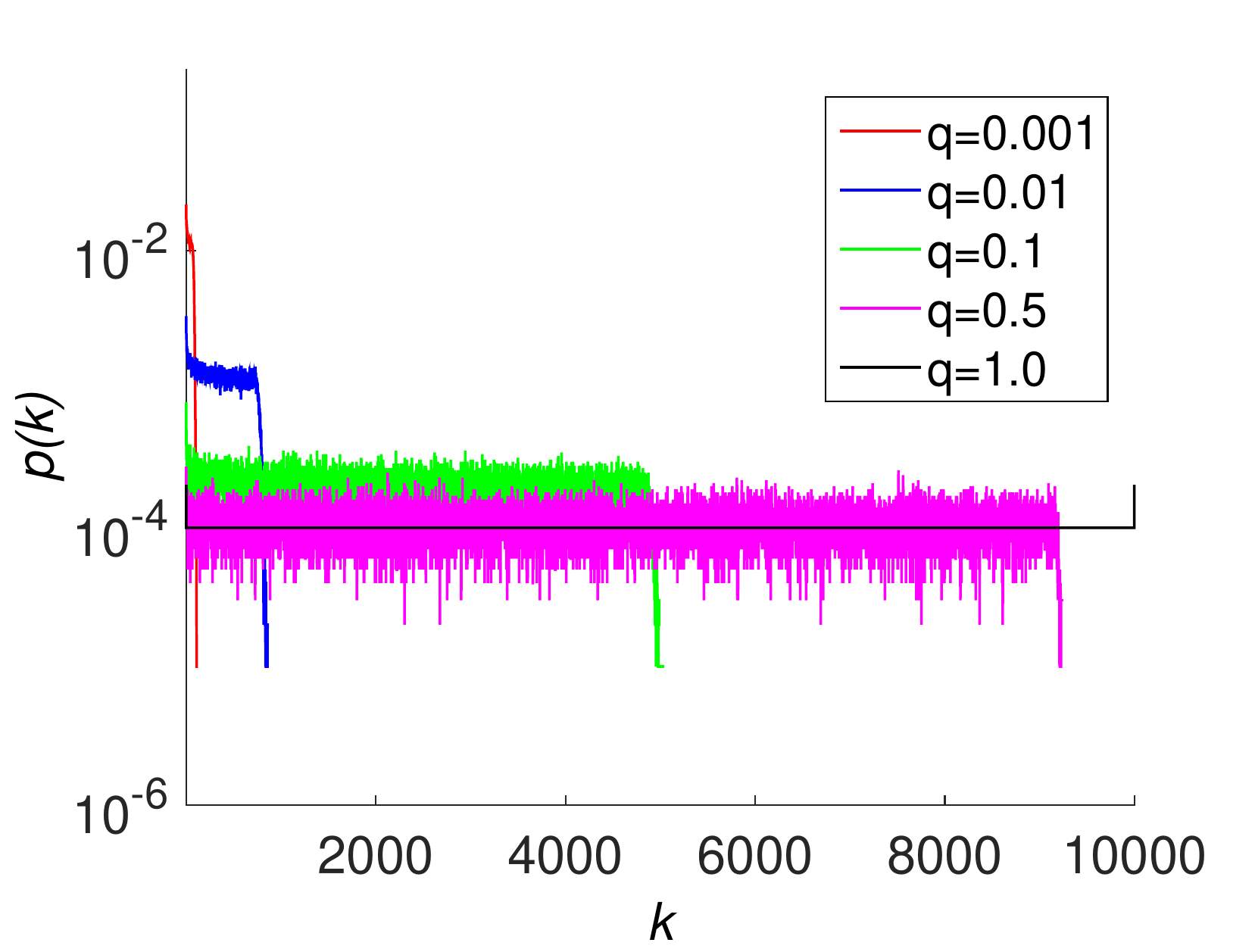}
\caption{Degree distributions when \textit{q} is 0.001, 0.01, 0.1, 0.5 and 1.0, respectively.}\label{fig7_kvsq}
\end{center}
\end{figure}

\subsection{Distribution of 4-motifs}

\noindent Motifs contribute to and even determine many basic properties of a network, thereby becoming
an important object for investigation. The distribution of the 4-motifs in the multiplex \textit{q}-snapback
network is shown in Fig.~\ref{fig8_motif}. There are 8 of 4-motifs as shown in Fig.~\ref{fig8_motif} (a),
labeled from A to H, respectively. Motifs in other sizes are either trivial or too complicated therefore are
not discussed here. Fig.~\ref{fig8_motif} (b) shows the distribution of each motif on the network $G$($q=0.1$,
$N=10,000$). As can be seen from the bar chat, the chains (motif type A) are the most frequently appearing
motif, followed by the loops (motif type D). In the next section, it will be shown that these two particular
4-motifs play key roles in the robustness of the network controllability.\\

\begin{figure}[ht!]
\begin{center}
\includegraphics[width=3.3in]{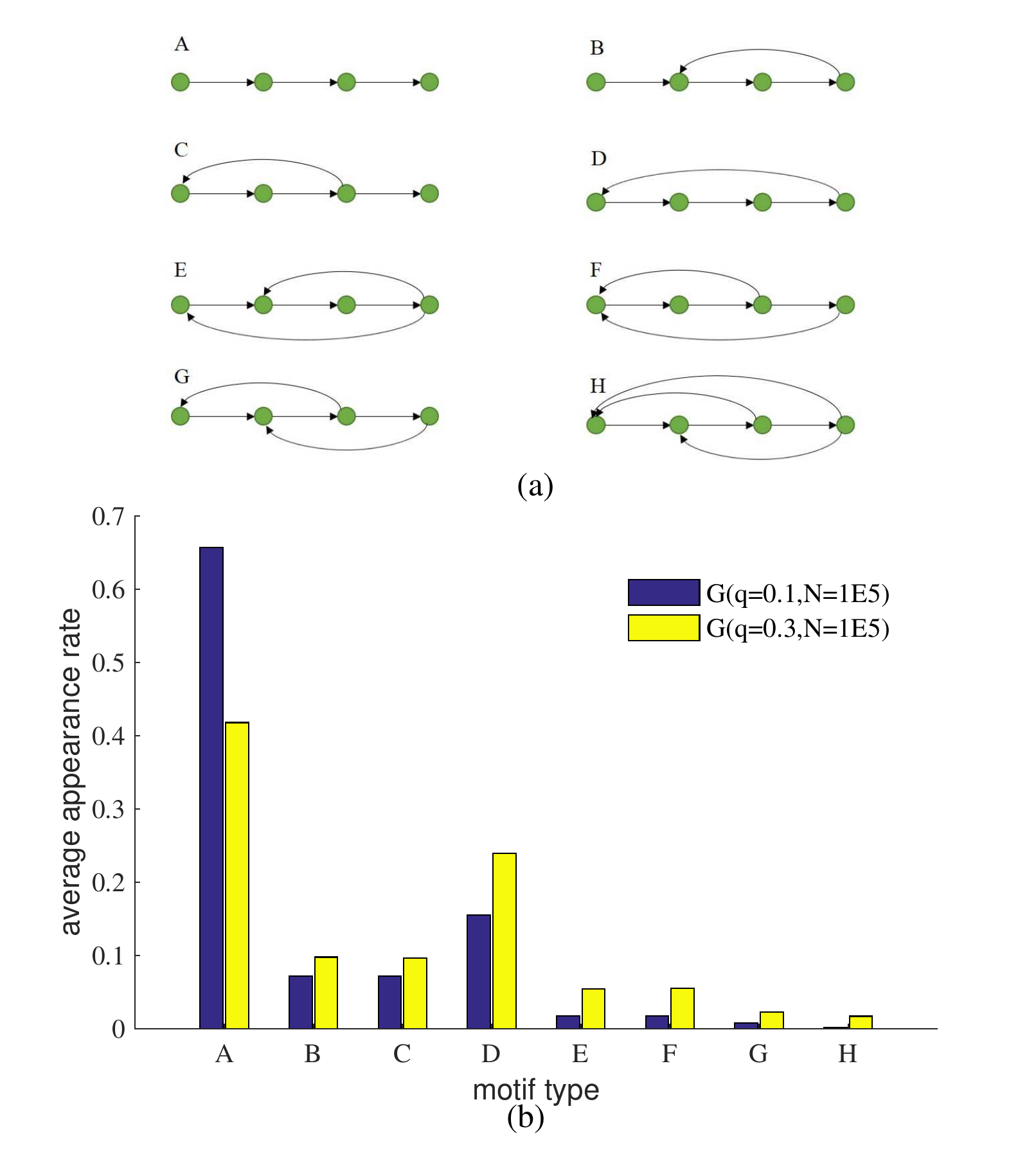}
\caption{Distribution of the 4-motifs: (a) notation of the eight 4-motifs;
(b) comparison of distributions of the eight 4-motifs on $G(q=0.1,N=10,000)$
and $G(q=0.3,N=10,000)$.
}\label{fig8_motif}
\end{center}
\end{figure}

\section {Controllability}\label{sec:ctrl}

\noindent The controllability of the \textit{q}-snapback network is studied through extensive simulations.

\begin{figure*}[htbp]
\begin{center}
\includegraphics[width=6in]{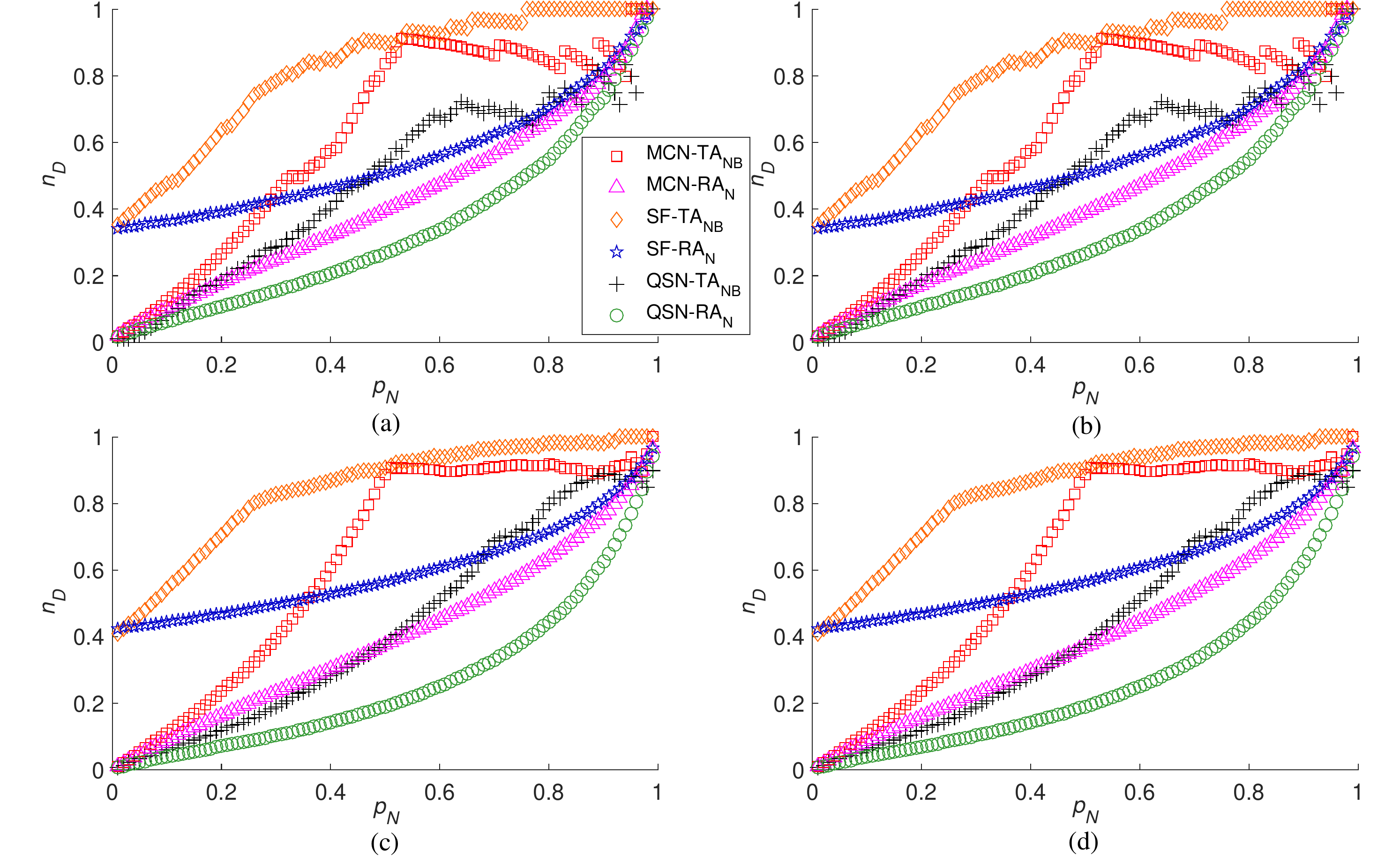}
\caption{Density of control-nodes $n_D$ as a function of the proportion $p_N$ of removed nodes,
where $n_D$ represents the proportion of needed control-nodes over all nodes of the current network.
TA\textsubscript{NB} represents the targeted attacks that aim at removing the node with the largest
betweenness on the current network: (a) Network size 100, state controllability; (b) network size 100,
structural controllability; (c) network size 1,000, state controllability; and (d) Network size 1,000,
structural controllability.}\label{fig9_nb}
\end{center}
\end{figure*}

Recall \cite{Wang2016A} that a system or network described by $\dot{{\bf x}}=A{\bf x}+B{\bf u}$,
where $A$ and $B$ are constant matrices of compatible dimensions, is {\it state controllable\/} if and
only if the controllability matrix $[B\ AB\ A^2B\ \cdots, A^{n-1}B]$ has a full row-rank, where $n$ is
the dimension of $A$. The concept of {\it structural controllability\/} is a slight generalization,
dealing with two parameterized matrices $A$ and $B$, in which the parameters characterize the structure
of the underlying system or network. If there are specific parameter values that can make the two
parameterized matrices become state controllable, then the underlying system or network is structurally
stable.

The network controllability is measured by the density of the control-nodes $n_D$, where $n_D\equiv
N_D/N$ and $N_D$ is the number of external controllers (also called driver nodes) needed to retain
the network controllability after the network had been attacked, and $N$ is the network size. The
smaller the $n_D$ is, the more robust the network controllability will be.

\begin{table}
\centering
\caption{Attack methods for simulation.}
\label{tab:tab1}
\begin{tabular}{|c|c|c|}
\hline
\multicolumn{1}{|l|}{} & \textbf{Node-removal} & \textbf{Edge-removal} \\ \hline
\textbf{Targeted} & \multicolumn{1}{l|}{\begin{tabular}[c]{@{}l@{}}TA\textsubscript{NB}:
to remove the node \\ with the largest betweenness\\ -----------------------------------\\
TA\textsubscript{ND}: to remove the node \\
with the largest degree\end{tabular}} & TA\textsubscript{E} \\ \hline
\textbf{Random} & RA\textsubscript{N} & RA\textsubscript{E} \\ \hline
\end{tabular}
\end{table}

The simulation design here is an extension of the multiplex congruence networks (MCN) simulations
reported in \cite{Yan2016SR}. Both MCN and scale-free (SF) networks are taken to compare with
the \textit{q}-snapback network. Scaling property is examined by using two network sizes, with
100 nodes and 1,000 nodes respectively. Five types of attacks, as shown in Table~\ref{tab:tab1},
are implemented, \textit{i.e.}, node-betweenness-based targeted attacks (TA\textsubscript{NB}),
node-degree-based targeted attacks (TA\textsubscript{ND}), node-based random attacks
(RA\textsubscript{N}), edge-based targeted attacks (TA\textsubscript{E}), and edge-based random
attacks (RA\textsubscript{E}). Here, TA\textsubscript{E} aims at removing the edges with the largest
edge-betweenness. To reduce the effect of randomness, the results of node-based RA are averaged
over 100 independent runs, and that of edge-based RA are averaged over 30 independent runs. The
detailed simulation results are shown in Figs.~\ref{fig9_nb},~\ref{fig10_nd}~and~\ref{fig11_e}.

\begin{figure*}[htbp]
\begin{center}
\includegraphics[width=6in]{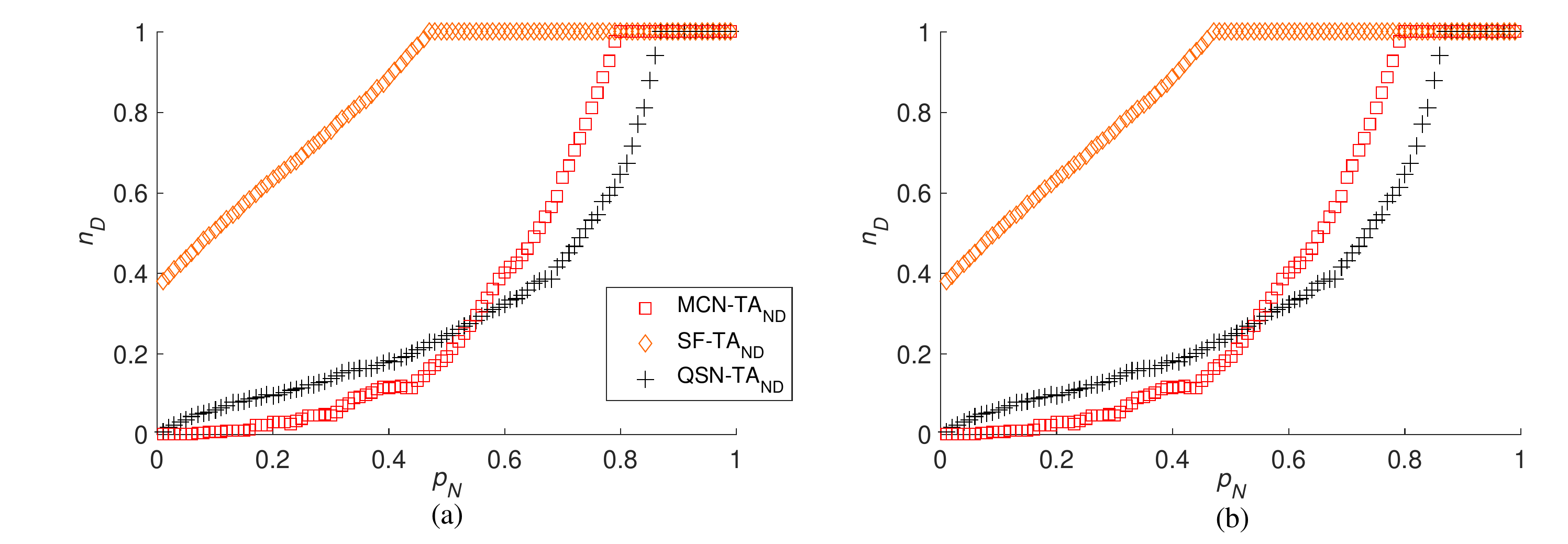}
\caption{Density of control-nodes $n_D$ as a function of the proportion $p_N$ of removed nodes.
TA\textsubscript{ND} represents the targeted attacks that aim at removing the node with the
largest out-degree in the current network, where for same-degree nodes it randomly removes one.
RA\textsubscript{N} represents the random attacks that randomly remove a node from the current
network: (a) Network size 1,000, state controllability; and (b) network size 1,000, structural
controllability.}\label{fig10_nd}
\end{center}
\end{figure*}

\begin{figure*}[htbp]
\begin{center}
\includegraphics[width=6in]{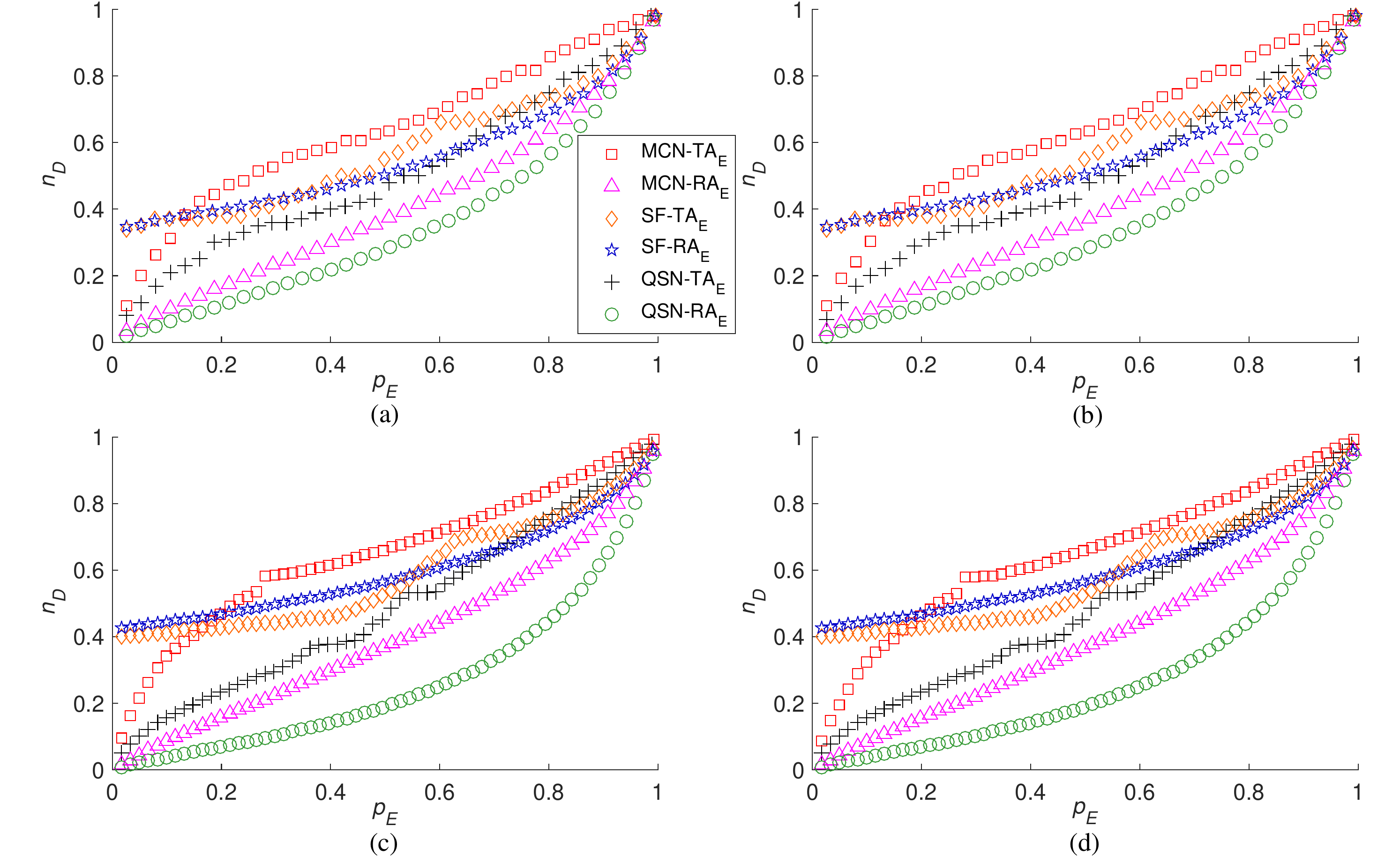}
\caption{Density of control-nodes $n_D$ as a function of the proportion $p_E$ of removed edges.
The subscript E represents that both the targeted and random attacks aim at removing edges:
(a) Network size 100, state controllability; (b) network size 100, structural controllability;
(c) network size 1,000, state controllability; and (d) Network size 1,000, structural
controllability.}\label{fig11_e}
\end{center}
\end{figure*}

When the network size is set to $N=100$, as did in \cite{Yan2016SR}, for comparison, the results
are shown in Figs.~\ref{fig9_nb} (a) and (b). Because the degree distribution is deterministic for
the MCN, in the case of 100 nodes it is $\langle k\rangle=3.82$ \cite{Yan2016SR}, the average
degrees of the other two types of (SF and \textit{q}-snapback) networks are set to $\langle k\rangle
\approx 3.82$ for a fair comparison, which means that they all have about the same number of links.
For SF networks, the average degree cannot be precisely controlled due to the randomness in their
generating processes, therefore fine-tunings are performed by adding or deleting a few links, so as
to slightly change the average degree such that the difference of average degrees between SF and MCN
networks becomes negligible. As for the \textit{q}-snapback network,
it turns out that $q=0.06$ is the best value to use, so
$G_2$($q=0.06$, $N=100$) is generated, which yields an average degree $\langle k\rangle=3.78$.
As a result, \textit{q}-snapback network is slightly sparsely linked as compared to MCN and SF.

When the network size is set to $N=1,000$, the average degree for the three types of networks is
set to $\langle k\rangle \approx 6.06$. Similarly, MCN has $\langle k\rangle=6.06$, SF network has
$\langle k\rangle \approx 6.06$ and the \textit{q}-snapback has $\langle k\rangle=6.055$; again,
the \textit{q}-snapback network is slightly sparser in terms of edge density in this case.

Both state controllability and structural controllability of these networks are examined.
Figs.~\ref{fig9_nb} (a) and (c) show a comparison on the state controllability under
TA\textsubscript{NB} and RA\textsubscript{N}, respectively. Both figures show the same phenomenon
regardless of the changes of the network sizes. The SF network is the most vulnerable to both
TA\textsubscript{NB} and RA\textsubscript{N}. The \textit{q}-snapback network is the most robust
against these two types of attacks. Likewise, in the structural controllability comparison as
shown in Figs.~\ref{fig9_nb} (b) and (d), the \textit{q}-snapback network is the most robust against
both TA\textsubscript{NB} and RA\textsubscript{N}.

When the target of the targeted-attacks is shifted, from node-betweenness (TA\textsubscript{NB})
to node-degree (TA\textsubscript{ND}), the \textit{q}-snapback network performs similarly to MCN.
More precisely, as shown in Fig.~\ref{fig10_nd}, MCN performs more robustly than \textit{q}-snapback
when $p_N<0.544$ ($p_N=0.544$ is the intersection of the curves MCN-TA\textsubscript{ND} and
QSN-TA\textsubscript{ND} in Fig.~\ref{fig10_nd}). This is because, by the network
construction and node-removal mechanisms, the node-degrees of the MCN are arranged in decreasing
order and the earlier node-based removals would not affect its connectivity, but after some time its
connectivity crashes rapidly. On the contrary, the node-degrees of the \textit{q}-snapback network
are uniformly distributed, so its connectivity remains about the same against removals even after
the MCN became disconnected.

Both MCN and the \textit{q}-snapback network outperform the SF network against the three types of
node-based attacks, essentially due to their inherent chain- and loop-motif structures.

Furthermore, simulation on an attack was performed on edge-removals, with results shown in
Fig.~\ref{fig11_e}. This kind of attack removes edges from the network, one after another, either
in the targeting order or at random. Note that, when the network size is 1,000, there are more
than six thousand edges in each network ($\langle k\rangle \approx 6.06$), thus the results of
random edge-removal are averaged over 30 independent runs. In this comparison, again,
the \textit{q}-snapback outperforms MCN and SF prominently.\\

\section{Conclusions}\label{sec:con}

\noindent A new complex network model, named \textit{q}-snapback network, has been introduced.
Some basic topological characteristics of the network have been calculated, including the degree
distribution, average path length, clustering coefficient and Pearson correlation coefficient.
The typical 4-motifs of the network have also been evaluated. Most importantly, the robustness
of both state controllability and structural controllability of the \textit{q}-snapback network
against five types of attacks ({\it i.e.}, targeted betweenness-based node-removal, targeted
degree-based node-removal, random node-removal, targeted edge-removal, and random edge-removal)
have been simulated with comparisons to the multiple congruence network and the generic scale-free
network, showing that the multiplex \textit{q}-snapback network has the strongest robustness of
both controllabilities due to its rich inherent chain- and loop-motif structure. The finding
reveals that, to build a network with strong robustness of controllabilities against node- and/or
-edge removal attacks, it is advantageous to embed more chain- especially loop-microstructures.
Whether or not such networks are also good in data traffic management, multi-agent systems and
industrial assembly-line automation, as well as other networking performances, remains an
important topic for future investigation.

Like the MCN, the $q$-snapback network has a backbone chain, which has advantage
in robustness but also has disadvantage as being less structurally flexible in modeling diverse
real-world networks. Looking forward, it is foreseeable that the new $q$-snapback network model
not only have potential applications in industrial assembly-line automation (Fig. \ref{fig1_eg}), but also
in biological systems \cite{Vinayagam2016PNAS} and brain science \cite{Tang2017AXV} regarding
its feedback control, controllability and information transmission.\\

\def\bibindent{1em}

\end{document}